\documentclass{article}

\usepackage{microtype}
\usepackage{graphicx}
\usepackage{subfigure}
\usepackage{booktabs} % for 
\usepackage{amsfonts}       
\usepackage{hyperref}

\usepackage{amsmath,array}

\usepackage[accepted]{icml2021}

\icmltitlerunning{Recovering AES Keys with a Deep Cold Boot Attack}

\begin{document}

\twocolumn[
\icmltitle{Recovering AES Keys with a Deep Cold Boot Attack}

\icmlsetsymbol{equal}{*}

\begin{icmlauthorlist}
\icmlauthor{Itamar Zimerman}{equal,TAU}
\icmlauthor{Eliya Nachmani}{equal,TAU,FB}
\icmlauthor{Lior Wolf}{TAU}
\end{icmlauthorlist}

\icmlaffiliation{TAU}{Tel Aviv University}
\icmlaffiliation{FB}{Facebook AI Research}

\icmlcorrespondingauthor{Itamar Zimerman}{zimerman1@mail.tau.ac.il}
\icmlcorrespondingauthor{Eliya Nachmani}{enk100@gmail.com}
\icmlcorrespondingauthor{Lior Wolf}{liorwolf@gmail.com}
\icmlkeywords{Machine Learning, ICML, Cold Boot Attack, Neural S-box, Side channel attack, Cryptanalysis}
\vskip 0.3in
]

\printAffiliationsAndNotice{\icmlEqualContribution} 

\begin{abstract}
Cold boot attacks inspect the corrupted random access memory soon after the power has been shut down. While most of the bits have been corrupted, many bits, at random locations, have not. Since the keys in many encryption schemes are being expanded in memory into longer keys with fixed redundancies, the keys can often be restored. In this work, we combine a novel cryptographic variant of a deep error correcting code technique with a modified SAT solver scheme to apply the attack on AES keys. 
Even though AES consists of Rijndael S-box elements, that are specifically designed to be resistant to linear and differential cryptanalysis, our method provides a novel formalization of the AES key scheduling as a computational graph, which is implemented by a neural message passing network. Our results show that our methods outperform the state of the art attack methods by a very large margin. 
\end{abstract}

\section{Introduction}
Many cipher architectures use expanded keys in their code. For reasons of efficiency, these algorithms do not re-calculate the expanded key with the expansion function each time a message is to be encrypted or decrypted. Instead, it is written to some RAM device (such as DDR and SRAM), until it is used. These devices often induce a security leak: after the power is shut down (even if the device is damaged, deleted, or burned) parts of the data still exist in the memory. This phenomenon is called data remanence \cite{data_remance_semicondcutr,Data_remanence_flash}. Its occurrence was established for a variety of memory devices \cite{DDR_and_MODERN,DDR3_REM}.  

A cold boot attack is a side-channel attack \cite{halderman_cba}, in which the attacker tries to recover the encryption key by exploiting the memory leakage and the redundancy of the key expansion function used by the encryption method.

This attack is well-known, and defense methods for it have been heavily researched \cite{DefCBA}. Vulnerable devices include computers and smartphones \cite{CBT_on_android,CBT_on_cellphones,CBT_on_cellphones2,CBT_on_laptop1,CBT_on_laptop2}. Examples of encryption systems that have been broken by this attack include Microsoft’s BitLocker, Apple’s FileVault, Linux’s Dm-crypt, TrueCrypt, Google’s Android’s disk encryption, and many others.

The data remanence phenomena are often modeled in one of two ways: a theoretical model and a more practical one. In both models, there is a tendency of decaying the bit values to the ground state, which can be $0$ or $1$. For simplicity, we assume that $0$ is the ground state. The strength of this trend depends on the hardware, the temperature, and the time that has elapsed since the power was turned off.

This model is formalized in the literature by the probability $\delta_0$ that a key-bit with a value of 1 on the original key will corrupt to value 0. Common values in the literature are in the range of $\delta_0 \in [0.3,0.75]$. In the theoretical model, we assume that no bit will corrupt from $0$ to $1$. The realistic model accounts for some reading errors for bits with an original value of $0$. Let $\delta_1$ be the probability that a key-bit with an original value of $0$ will corrupt to value $1$. In the theoretical model $\delta_1 = 0.0$, and in the realistic model $\delta_1 \in \{0.0005,0.001\}$.

The main computational problem is the one of recovering an encryption key from its corrupted key by using the redundancy that is induced by the key expansion function. In this paper, we present a new algorithmic method based on deep learning for the key recovering problem. While the previous leading methods are based completely on SAT solvers, our method contains a deep network that provides an estimate of the solution, which is subsequently used to seed the SAT solver. The network employs techniques from the field of error-correcting codes, which are further enhanced by cryptographic components that we term as neural S-boxes. Our method is employed in order to drastically improve an existing algorithm for recovering AES-256 keys \cite{AES}. AES is the most popular symmetric cipher and it considered to be completely secure. 

\section{Related Work}

We discuss cold boot attacks and then neural networks for error correcting codes. The latter provides a solid foundation for our technique.

\subsection{Cold Boot Attack}

The first algorithm invented for cold boot attacks was by \citet{halderman_cba}. Using techniques from the field of error correcting codes, a method was presented for recovering DES, AES, tweak and RSA keys. Subsequent methods were based on a variety of techniques and ideas, such as integer programming and solving polynomial systems of equations \cite{CBaPoliynomialSYS}, SAT solvers \cite{SATforCBA} , MAX-SAT solvers \cite{Liaobbb} and more \cite{tsow2009improved}, \cite{tanigaki2015maximum}.

Our work focuses on AES-$256$, for which we are aware of two existing contributions. The work of ~\citet{tsow2009improved} recovers AES key schedules from decayed memory images. The TSOW algorithm was presented for the theoretical model, and the result was a success rate of $99.4\%$ for $65\%$ decay rates. However for $70\%$ the success rate was $0.0$. \citet{tanigaki2015maximum} presented an algorithm for the realistic model, which is based on a combination of the TSOW Algorithm and maximum likelihood approach. A theoretical analysis showed that $\delta_0$ must be smaller than 0.75, when $\delta1 = 0.001$. We note that our success rate is not negligible in this setting and compare with the method extensively in our experiments.

There are considerably more contributions for the AES-128 cipher, which presents an easier problem since the search space is reduced drastically with smaller keys. \citet{halderman_cba} had a 70 $\%$ success rate for $\delta_0 = 0.15 $ and $\delta_1 = 0.001$. This result was improved by \citet{tsow2009improved} with the TSOW algorithm. In the work of \citet{SATforCBA} and \citet{Liaobbb}, two SAT-based algorithms were proposed. 

The first algorithm was designed for the theoretical decay model, and achieved a success rate of 100$\%$ for $\delta_0=0.78\%$. However for $80\%$ the results were limited. The second algorithm was designed for the realistic model, and is based on a partial MAX-SAT solver. The success rate of the algorithm was $100\%$ for $\delta_0 = 0.76$ and $\delta_1 = 0.001$. 
Due to this success on the smaller keys, we implemented both the SAT solver and the MAX-SAT solver techniques for the aes-256 keys, and compare the results to our method.

\subsection{Error Correcting Codes with Deep Learning}

Deep learning was applied to various error correcting codes over the past few years. Polar codes \cite{tal2013construct} which are use in 5G cellular communication, can be decoded with neural networks with neural successive cancellation decoding \cite{doan2018neural,gross2020deep}. Moreover, an improved deep Polar decoding is introduced in \cite{gruber2017deep,xu2017improved,teng2019low}. 

In \cite{an2020high}, the Reed-Solomon neural decoder is introduced, which estimates the error of the received codewords, and adjust itself to do more accurate decoding. Neural Bose–Chaudhuri–Hocquenghem (BCH) codes decoding is introduced in \cite{kamassury2020iterative, nachmani2019hyper,raviv2020data}. 

Low-Density-Parity-Check (LDPC) neural decoding is introduced in \cite{habib2020learning}. The paper demonstrates a novel method for sequential update policy in the Tanner graph. In \cite{jiang2019turbo} a deep Turbo autoencoder is introduced for point-to-point communication channels. Furthermore, \cite{kim2018deepcode,kim2018communication} present a novel method for designing new error correcting codes by neural networks. In \cite{caciularu2020unsupervised} a neural channel equalization and decoding using variational autoencoders is introduced. A deep soft interference cancellation for MIMO Detection are present in \cite{shlezinger2020deepsic}.

In this work, we will focus on neural belief propagation decoding, as described by \citet{nachmani2016learning}. This work demonstrated that short block codes, with up to a few thousand bits, can be better decoded by a neural network than by the vanilla belief propagation algorithm that the network is based on. This method is highly relevant for correcting AES corrupted keys and cold boot attacks, since the length of the AES expansion key has a few thousand bits. 

\section{Background}

\subsection{AES-256 Key Expansion Function} 
\label{section:keyExFunc}
The AES algorithm \cite{AES} is based on the key expansion function $f$, which operates on a random $256$-bit initial key
\begin{equation}
f : \{0,1\}^{256} \rightarrow  \{0,1\}^{1920}
\end{equation}

$f$ is computed in iterations, also known as rounds. In each iteration, $128$ bits of the expansion are calculated from the previous bits. The calculation consists of both linear and non-linear operations. The non-linear ones are called the Rijndeael substitution box or S-box for short.

The Rijndeael S-box function is described in detail in Chapter 4.2.1 of \cite{AES} and it is usually implemented as a look-up-table. It is composed of two transformations: (i) an affine transformation and (ii) Nyberg S-box transformation \cite{nyberg1991perfect}. The Nyberg S-box transformation is a mapping of an input vector to its multiplicative inverse on the Rijndael finite field: $x \rightarrow x^{-1}$ in $GF(2^8)$. This transformation is known as a perfect non-linear transformation and satisfies certain security criteria.

We will use the following notations:
\begin{enumerate}
    \item Denote the expanded key bits as $\hat{w} := (w_0, .. ,w_{n-1})$, where $n$ is the size of the expanded key, which is equal to $1920$. Denote the byte $w_i,...,w_{i+7}$ as $W_i$, and the double word $w_i,...,w_{i+31}$ as $W'_i$, so $W'_i(j) = w_{i+j}$.
    \item Let $S : \{0,1\}^8 \rightarrow \{0,1\}^8$ be a Rinjdeal S-box. We can extend the definition of $S$ to an input vector of $32$ bits, where the result is obtained by applying S on each byte separately.
    \item $c = c_1,...,c_{10}$  is a vector of fixed values that is defined in the RCON table which is given in \cite{AES}. This constant is used in the key expansion function $f$.
    \item $R$ is the following rotation function: 
    \begin{equation}
        R(w_1,...,w_7,w_8,...,w_{32}) =  (w_8,...,w_{32},w_1,...,w_7) 
    \end{equation}
    which is used in the key expansion function $f$. 
    \item $k$ is the initial key size $256$, and $b$ is the block size $128$.
    \item $ \% $ is the modulo operator and $\oplus$ is XOR operator.
    \item For each key index $i$, we denote the round number as $r(i) = \lfloor\frac{i}{b}\rfloor$, and the double word number as $d(i) = \lfloor\frac{i}{32}\rfloor$ 
    \item $\tau = \frac{n-k}{b}$ is the number of rounds in the key expansion function.
\end{enumerate}

The key expansion function is critical to understanding our method. Here we describe the constraints that this function inducts on the key bits. For the $i$-bit in the key, the constraints are given by:  
\begin{enumerate}
\item $ \forall i: k \leq i < n , i \% b < 32, r(i) \% 2 = 0:
$
\begin{equation}
\label{eq:Sbox_with_rotation}
w_i = w_{i - k} \oplus S(R(W'_{d(i-32)}))(i\%b) \oplus c_{\frac{r(i)}{2}}
\end{equation}

\item
$ \forall i: k \leq i < n  ,  i \% b < 32, r(i) \% 2 = 1:
$
\begin{equation}
\label{eq:Sbox_without_rotation}
w_i = w_{i - k} \oplus S(W'_{d(i-32)})(i\%b)
\end{equation}

\item
$ \forall i: k \leq i < n  , i \% b \geq 32 : $
\begin{equation}
\label{eq:XOR_equation}
w_i = w_{i - k} \oplus w_{i -32}
\end{equation}
 \end{enumerate}
Note that each equation contains three XOR operations between variables, and in some of the equations there is a XOR with a constant value.

\subsection{Error Correcting Codes with Deep Belief Propagation}
\label{subsection:BP}
A deep learning decoder for error correcting codes with a belief propagation algorithm was introduced in \cite{nachmani2016learning}. The decoding process uses the well-known belief propagation method and adds learnable weight to the algorithm. Specifically, they add weights to the edges in the Trellis graph. For a linear block code with $k$ information bits and $n$ output bits, the parity check matrix of the linear block code $H$ has a size of $(n-k) \times n$. 

The deep neural Belief propagation algorithm that was introduced in \cite{nachmani2016learning} has an input layer of $n$ bits. In the architecture that is defined in \cite{nachmani2016learning} there are two types of hidden layers which are interleaved: (i) variable layer for odd index layer $j$ and (ii) check layer for even index layer $j$.

For notational convenience, we assume that the parity check matrix $H$ is regular, meaning, the sum over each row and column is fixed and denoted by $d_v$ and $d_c$ respectively. Each column of the parity check matrix $H$ is corresponding to one bit of the codeword and obtains $d_v$ variable nodes in each variable layer. Therefore, the total number of \textit{variable processing units} in each variable layer is $E=d_v n$. Similarly, each check layer has $E=(n-k)\times d_c$  \textit{check processing units}.

During the decoding process, the messages propagate from the variable layer to the check layers iteratively, where the input to the network is the log likelihood ratio (LLR) $\ell \in \mathbb{R}^{n} $ of each bit:
\begin{equation}
\ell_v = \log\frac{\Pr\left(c_v=1 | y_v\right)}{\Pr\left(c_v=0 | y_v\right)},
\end{equation}
where $\ell_v$ is the log likelihood ratio for each received signal $y_v$ and $c_v$ is the bit that we want to recover.

Denote $x^j$ as the vector messages that propagate in the Trellis graph. For $j=1$ and for odd $j$, the computation in each variable node is:
\begin{equation}
\label{eq:odd}
x^{j}_e = x^{j}_{(c,v)} = \tanh \left(\frac{1}{2}\left(l_v + \sum_{e'\in N(v)\setminus \{(c,v)\}} w_{e'}x^{j-1}_{e'}\right)\right)
\end{equation}
where $N(v)=\{(c,v) | H(c,v)=1\}$ is the set of all edges that are connected to $v$ and each variable node indexed the edge $e=(c,v)$ in the Tanner graph. $w_e$ is a set of learnable weights.

For even layer $j$, each check layer performs  this computation:

\begin{equation}
\label{eq:even}
x^{j}_e = x^j_{(c,v)} = 2arctanh \left( \prod_{e'\in N(c) \setminus \{(c,v)\}}{x^{j-1}_{e'}}\right) 
\end{equation}
where for each row $c$ of the parity check matrix $H$, $N(c)=\{(c,v) | H(c,v)=1\}$ is the corresponding set of edges.

Overall, in the deep neural network that is proposed in \cite{nachmani2016learning} there are $L$ layers from each type (i.e. variable and check). The last layer is a marginalization layer with a sigmoid activation function which outputs $n$ bits. The $v$-th output bit is given by:

\begin{equation}
o_v = \sigma \left( l_v + \sum_{e'\in N(v)}\bar{w}_{e'} x^{2L}_{e'} \right),
\label{eq:base_final}
\end{equation}
where $\bar{w}_{e'}$ is another set of learnable weights. Moreover, in each odd layer $j$, marginalization is performed by:
\begin{equation}
o^{j}_v = \sigma \left( l_v + \sum_{e'\in N(v)}\bar{w}_{e'} x^{j}_{e'} \right)
\label{eq:j_final}
\end{equation}

The loss function is cross entropy on the error after each $j$ marginalization:
\begin{equation}
\mathcal{L}=-\frac{1}{n}\sum_{h=0}^{L}\sum_{v=1}^{n} c_{v}\log(o^{2h+1}_{v})+(1-c_{v})\log(1-o^{2h+1}_{v})
\label{eq:loss}
\end{equation}
where $c_{v}$ is the ground truth bit.
  
\begin{figure*}[t]
   \centering
       \includegraphics[width=.6\textwidth]{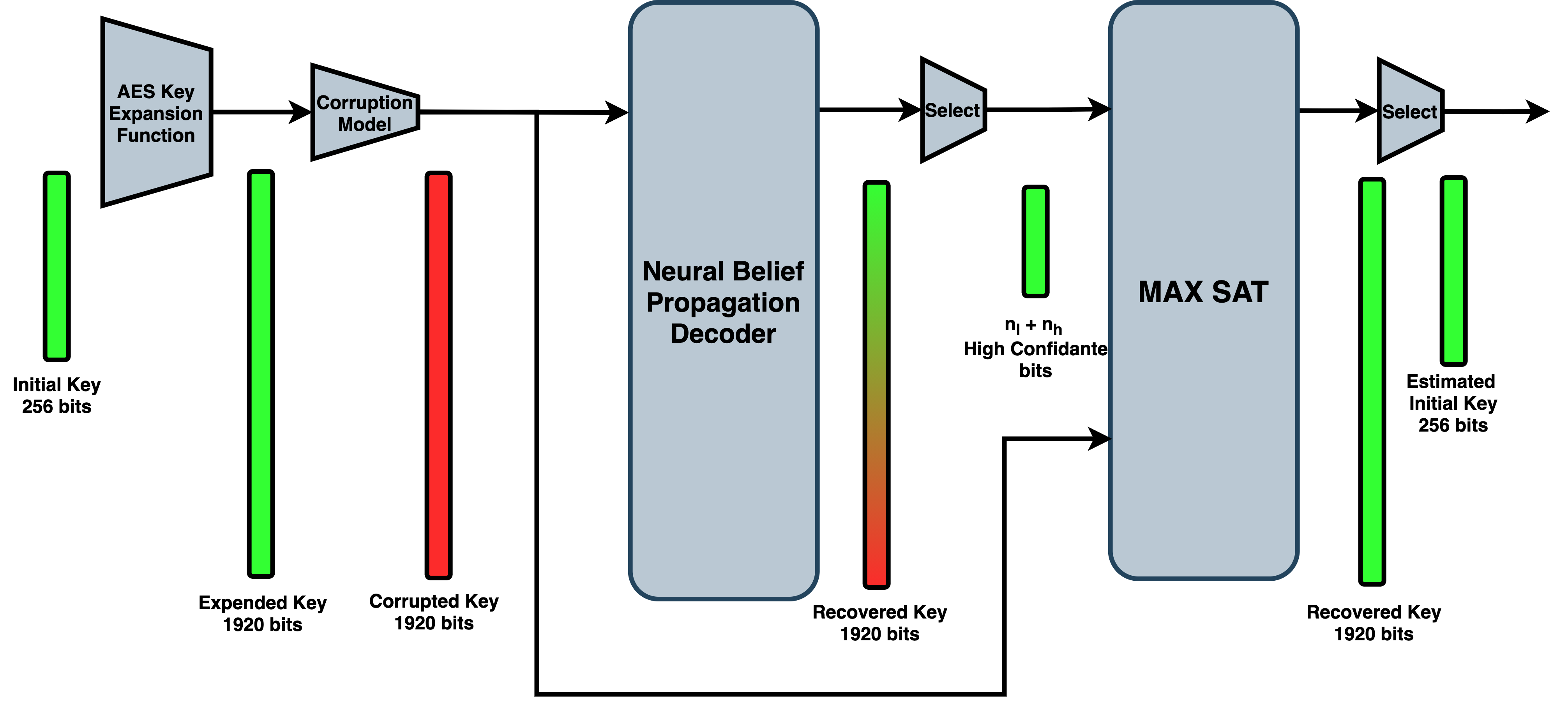}
    \caption{An overview of our method for a neural cold boot attack. The input is the initial key of $256$ bits, then the AES key expansion function $f$ expands it to $1920$ key. The expanded key was corrupted by the cold boot model. The corrupted key is inserted into a cryptographic neural belief propagation decoder whose constructs form a novel formalization of the AES key expansion function. The most accurate $n_l+n_h$ bits are then selected in insert with the corrupted key to MAX-SAT solver. The MAX-SAT solver produces the corrected AES key.}
 \label{fig:arch}
\end{figure*}

\section{Method}

Our architecture contains two components: (i) A variant of neural belief propagation decoder with neural S-box layers and (ii) a Partial MAX-SAT solver. The proposed model is depicted in Figure~\ref{fig:arch}. The input to the neural belief propagation is the corrupted bits $l = l_0,..,l_{n-1}$ and it predicts an approximation for the original key $o = o_0,..,o_{n-1}$. Formally, the value of the $i$-th bit in the original key was $1$ with an approximated probability of $o_i$. The input to the Partial MAX-SAT solver is a CNF formula with part of it defined by $o' \subset o$ , where the probabilities in $o'$ correspond to bits that the network has high confidence in their values (approximately 99$\%$). The output of the Partial MAX-SAT solver is the estimation of the desired key.

\subsection{Rijndael S-box as Neural Network and the S-box Layer}
\label{section:sbox_as_nn}
The belief propagation neural network is defined by a parity check matrix $H$. However, the AES constraints in Eq.~\ref{eq:Sbox_with_rotation},\ref{eq:Sbox_without_rotation} are non-linear, since they include the S-box transformation. Therefore, there is no parity check matrix $H$ such that $Hl=0$.

In order to convert these equations to linear form, one can change the variables by concatenating $W_t^{s'}:= S(W_t^{'})$ to $W_i$, and
construct a parity check matrix $H$ for the new variables. However, since the S-box transformation is defined by boolean vectors, and the neural belief propagation uses fractions values another problem arises, one cannot calculate $W_t^{s'}$ between layers of the neural belief propagation.

Therefore, in order to obtain a continuous and differentiable version of the Rijndael S-box, we first train a neural network $S_{nn}$ to mimic it:
\begin{equation}
    S_{nn} : x_1 , ... , x_8 \rightarrow y_1 , ... , y_{256}
\end{equation}
where $x_i \in [0,1]$ and $y_i \in [0,1]$. The network has three fully connected layers with $512$ ReLU activations. It is trained with a cross entropy loss function. An $argmax$ operation is performed on the output $y$, to obtain $z_1 , ... , z_{8}$, where it achieves $100\%$ accuracy. 

We can extend the definition of $S_{nn}$ to an input vector of 32 bits, where the result is obtained by applying $S_{nn}$ on each byte separately.

While the neural s-box is applied to bytes, the s-box layer is applied to the entire expanded key, which is constructed from a combination of neural S-boxes. Given an input vector $\hat{x} \in [-1,1]^{n}$, the s-box layer calculates the output vector $\hat{s} \in [-1,1]^{n+32 \tau+1}$ which is obtained by concatenation of the input with $32 \tau$ elements, which are calculated by applying the neural S-boxes on the corresponding bits of the input as follows:
\begin{enumerate}
    \item $\forall i \in [0,n-1] : \hat{s}_i = \hat{x}_i $
    \item $\forall i \in [n,(n+32\tau)-2] , i \% 32 =0:$ 
    
    $(\hat{s}_i, .., \hat{s}_{i+31})=S_{nn}(\hat{x}_{si(i)}, .., \hat{x}_{si(i)+31})$
    
    where $si(i)=b ((i-n)\%32) + k-32$
    \item The last bit in $\hat{s}$ is $1$ (namely $\hat{s}_{n+32\tau -1} = 1$) and it is used as a bias factor.
\end{enumerate}
There are $\tau$ additional rounds on the AES-$256$ key expansion, on each round $4$ S-boxes are calculated. Overall, each s-box layer consist of $4\tau$ neural S-box instances.

\subsection{Tailoring the Neural Belief Propagation}
To predict the values of the AES key from the corrupted keys, a naive way is to search the key that is close to the corrupted one over the large space of the existing key.  

However, when the decay percentage is high, the search space is extremely large, and by design, the key expansion function provides resistance against attacks in which part of the cipher key is known, see Chapter 7.5 in \cite{AES}.

Instead, due to the resemblance of the AES keys and the key role of the XOR operation, we rely on network-based error correcting code methods that are suitable for block ciphers. Such codes often employ expansion functions, which are, however, linear. 

\begin{figure*}[t!]
  \centering
      \includegraphics[width=.8\textwidth]{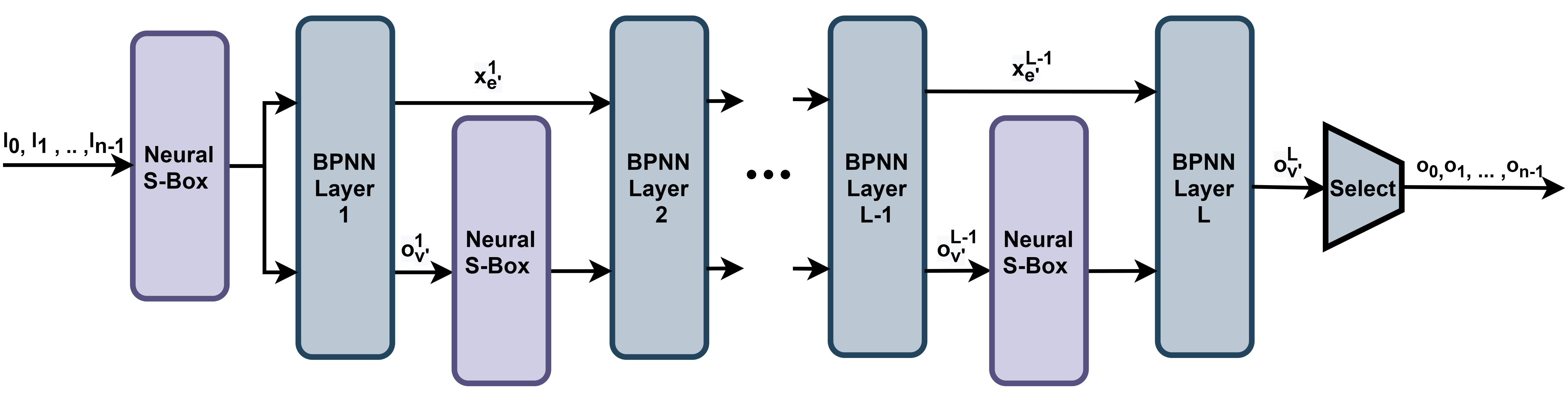}
    \caption{The architecture of the cryptographic neural belief propagation. The input is the corrupted key of the $1920$ bits. Each neural belief propagation layer receives two vectors: (i) the output of the previous belief propagation layer $x_{e'}^{j-1}$, (ii) the output of the marginalization layer $o_{v}^{j-1}$, after going through the S-box layer. After the last iteration, we cut bits whose not approximated bits in the corrupted key.}
 \label{fig:BPNN_with_Sboxs}
\end{figure*}

We modify the neural belief propagation architecture as follows: (i) adding a S-box layer after each check layer, (ii) modify the layer structure: replacing $\ell_v$ 

in Eq.~\ref{eq:odd},~\ref{eq:base_final},~\ref{eq:j_final} with the output of the marginalization layer $o_{v}^{j-1}$ on the previous iteration after it goes through the S-box layer. Figure \ref{fig:BPNN_with_Sboxs} we depict the architecture of the modified neural belief propagation.

\label{subsection:Tailoring_BPNN}

\subsection{Defining the ECC constraints}
\label{subsection:ECC_constraints}

Denote the S-box mimicking network output, given by a vector $W_i$, as $Z_i = (z_i, .. ,z_{i+7}) := S_{nn}(W_i)$. We denote the concatenation of $ Z_i,Z_{i+8},Z_{i+16},Z_{i+24}$ as $\hat{Z_i}$. 

We can rearrange the constraints in Eq.~\ref{eq:Sbox_with_rotation},~\ref{eq:Sbox_without_rotation},~\ref{eq:XOR_equation} as follows:  
\begin{enumerate}
\item $ \forall i: k \leq i < n , i \% b < 32, r(i) \% 2 = 0:$
\begin{equation}
\label{eq:NN_With_rotations}
0 = w_i \oplus w_{i - k} \oplus S_{nn}(R(W^{'}_{d(i-32)}))(i\%b) \oplus c_{\frac{r(i)}{2}}
\end{equation}

\item
$ \forall i: k \leq i < n  ,  i \% b < 32, r(i) \% 2 = 1:
$
\begin{equation}
\label{eq:eq_bb}
0 = w_i \oplus w_{i - k} \oplus S_{nn}(W^{'}_{d(i-32)})(i\%b)
\end{equation}

\item
$ \forall i: k \leq i < n  , i \% b \geq 32 : $
\begin{equation}
\label{eq:sim_xor}
0 = w_i \oplus w_{i - k} \oplus w_{i -32}
\end{equation}
 
\end{enumerate}

We define $x$ as the concatenate of $w$ and $\hat{Z}_i$: $x = (w_0,..,w_{n-1},\hat{Z}_{k-32},\hat{Z}_{k-32+b},\hat{Z}_{k-32+2b},..,\hat{Z}_{n-32+\tau b})$.
By considering the XOR operation as the addition operator over $\{0,1\}^2$, assuming for simplicity that $R$ is the identity function, so $S_{nn}(R(\hat{W}_{d(i-32)})) = \hat{Z_i}$ and replacing $S_{nn}(W_{i}')$ with $\hat{Z_i}$, one can transform Eq.~\ref{eq:NN_With_rotations}, ~\ref{eq:eq_bb}, ~\ref{eq:sim_xor} to a matrix form using a matrix $H'$ and a vector $u$, such that:
\begin{equation}
\label{eq:H_with_bias}
H'x+u=0
\end{equation}
where $u$ is a constant vector that consists of the RCON values $c_i$ and zeros %.
and $H'$ is a matrix with $n-k$ rows, as the number of bits calculated by the expansion function, and $n+32\tau$ columns, as the number of variables.

$\forall i,j :  0\leq i < n-k, 0\leq j < (n+32\tau)$

\begin{equation}
\label{eq:define_H}
H'(i,j)= \begin{cases}
			1, \text{if $i \% b \geq 32$ , $j = i$ }\\
			1, \text{if $i \% b \geq 32$ , $j = i+k$ }\\
			1, \text{if $i \% b \geq 32$ , $j = i+k-32$ }\\
            1, \text{if $i \% b < 32 $ , $j = i$ }\\ 
            1, \text{if $i \% b < 32 $ , $j = i+k$ }\\ 
            1, \text{if $i \% b < 32 $ , $j=n + 32 r(i) +i\%32$ }\\ 
        
            0, \text{otherwise}\\
		 \end{cases}
\end{equation}

The first three cases correspond to Eq.~\ref{eq:sim_xor}, the following three cases correspond to Eq.~\ref{eq:NN_With_rotations} ,~ \ref{eq:eq_bb}.

Moreover, $u$ is the constant vector that consists of the RCON values and defined by
$\forall i :  0\leq i<n-k: $
\begin{align*}
\label{eq:define_b}
 u_i = \begin{cases}
            c_{ \frac{r(i)}{2} },&\text{if $i \% b < 32, r(i) \% 2=0$}\\
            0, & \text{otherwise}\\
		 \end{cases}
\end{align*}
%\end{equation}
Note that without assuming that $R$ (in Eq.~\ref{eq:NN_With_rotations}) is the identity function, rather than a rotation function, one can rewrite the same formulation with a single difference, applying permutation on the vector $x$, or modify the equations where $r(i)$ is even. We did not use this assumption in practice (implementation and experiments).

It remains to convert Eq.~\ref{eq:H_with_bias} to a homogeneous form, by using the bias trick. Instead of the XOR operation with the bias $u$ in Eq.~\ref{eq:H_with_bias}, concatenate one bit with a constant value of $1$ to $x$. This bit used as a bias factor, and by using $H$, a concatenate of $H'$ with $u$, we can formulate Eq.~\ref{eq:H_with_bias} as follows:

\begin{equation}
\centering
\label{eq:H_mul}
H[x,1] = 
H
\begin{bmatrix}
           w_{0} \\
          % w_{2} \\
           \vdots \\
           w_{n-1} \\
           \hat{Z_{k-32}} \\
           \hat{Z_{k-32+b}} \\
           \hat{Z_{k-32+2b}} \\
           \vdots \\
           \hat{Z_{n-b-32}} \\
           1
         \end{bmatrix} = 0
\end{equation}
$\forall i,j :  0\leq i < n-k, 0\leq j \leq(n+32\tau): $

\begin{equation}
\label{eq:define_H}
H(i,j)=\begin{cases}
            c_{ \frac{r(i)}{2} }, \text{ if } j = (n+32\tau) \text{ and } \\ i \% b < 32  \text{ and } r(i)\% 2 =0\\
			1, \text{if $i \% b \geq 32$ , $j = i$ }\\
			1, \text{if $i \% b \geq 32$ , $j = i+k$ }\\
			1, \text{if $i \% b \geq 32$ , $j = i+k-32$ }\\
            1, \text{if $i \% b < 32$ , $j = i$ }\\ 
            1, \text{if $i \% b < 32$ , $j = i+k$ }\\ 
            1, \text{if $i \% b < 32$ , $j = n + 32 r(i) +i\%32$ }\\ 
            0,  \text{otherwise}\\
		 \end{cases}
\end{equation}

Note that the formulation of $H$ also relevant for other variations of AES (i.e. $k$=$128$,$192$). Moreover, the same technique can be used to create deep architectures for side-channel attacks for additional ciphers, for example Serpent ~\cite{SERPENT}.

Based on the $H$ matrix described in Eq,~\ref{eq:define_H}, we construct a neural belief propagation network, as described in Sec.~\ref{subsection:BP}. 

\subsection{Partial MAX-SAT Solver} 
\label{section:SAT_SOLVER}

Once we obtain the initial estimate from the neural network, we use a Partial MAX-SAT Solver to search for the corrected key. To run the solver, we define the following Conjunctive Normal Form (CNF) formulas:

\begin{enumerate}
    \item n variables, one per bit in the key $v_1 ,..,v_n$.
    \item Converted the bit-relation in Eq.~\ref{eq:Sbox_with_rotation},\ref{eq:Sbox_without_rotation},\ref{eq:XOR_equation} that implies by the key expansion function to a CNF formula by CNF Factorization. The result is the formula $\psi_{AES}$, that consists of $217984$ clauses and $1920$ variables.
    Eq.~\ref{eq:XOR_equation} for example, which is in the following form: ($ a \oplus b = c$), is replaced with the following clauses:
    $$
    (i)\quad \neg a \land \neg b \land \neg c \quad \quad \quad (ii)\quad \neg a \land b \land c 
    $$
    $$
    (iii)\quad a \land \neg b \land c \quad\quad\quad
   (iv) \quad a\land b \land \neg c
    $$

   With the other equations, the result is more complicated, and each equation has been replaced by numerous clauses. We then insert these clauses into the solver as a hard formula. This formula is identical for all of the instances and is calculated once in pre-processing.
   
   \item For each bit whose value is $1$ in the corrupted key, we supply a single clause that enforces this key bit to be $1$, we denote this formula by $\psi_{memory}$.
   Formally: 
   
   $ \psi_{memory} :=   \wedge_{i \in [n-1], l_i = 1 } v_i  $
   \item Consider the $n_h$ bits with the highest value in the network output $o$, and the $n_l$ bits with the lowest values. These are the locations for which the network is mostly confident. Let $t_h$ be the $n_h$-th highest value in $o$, and the $t_l$ as the $n_l$-th lowest values in $o$, we take these as thresholds and  define the formula: 
   
   $ \psi_{nn} :=   \Big{(}\wedge_{i \in [n-1], o_i \geq t_h}\neg v_i \Big{)}\wedge \Big{(}  \wedge_{i \in [n-1], o_i \leq t_l } v_i \Big{)}$
\end{enumerate}
We define $ \psi_{AES} $ as hard formula, and $\psi_{nn}$ as soft formula. 
In the theoretical decay model, $\psi_{memory}$ is defined as hard formula, however in the realistic decay model is defined as a soft formula.

There is a large number of Partial MAX-SAT Solvers, which operate in a wide variety of strategies. We select the WBO Solver~\cite{wbo_solver} with the implementation of ~\cite{LogicNGgithub}, which is based on the unsatisfiability method \cite{UNSAT}, and other enhancements \cite{WBO1,WBO2,WBO3}. We select this solver for three main reasons:
\begin{enumerate}
    \item We have the intuition that DPLL solvers will be suitable for this problem over randomized SAT solvers due to the large complexity of the search space (there are $2^{1920}$ vectors, and there are numerous clauses). This complexity makes us think that it is better to use a solver that scans the space in an orderly manner. We, therefore, decided to use CDCL solvers ~\cite{CDCL1,CDCL2,CDCL3}, the most popular variation of DPLL solvers.
    \item Since it achieved the best results in different cold boot attack settings, for example \cite{Liaobbb}.
    \item { In the early step of this development, we tried to insert the complete key approximation from the neural network into a CDCL solver, instead of using $\psi_{nn}$ and $\psi_{memory}$. Empirically, we observe that inserting the complete key approximation from the neural network into a CDCL solver does not output the correct keys.} Therefore, we decided to focus on a small number of bits. We chose the bits that the neural belief propagation is relatively sure in their values, and in total, the probability that more than a few bits in the subset are errors is very small (smaller than $1 \%$). Therefore, it was natural to use the UNSAT ~\cite{UNSAT} method which is suitable for the problem, since the number of unsatisfiability soft clauses is small with high probability.
    
\end{enumerate}

\subsection{The Overall Model}
The input of our architecture is the vector $l_1, ..,l_n$, which represents the corrupted key. It is inserted into a cryptographic variant of the neural belief propagation, which includes S-box layers. The S-box layers are based on the fully-connected neural network $S_{nn}$, which imitates the Rijndael S-box and extends its functionality to non-binary values. 

The original neural belief propagation layers are defined by a parity check matrix $H$, which designed according to the key expansion function, as designed in Eq.~\ref{eq:define_H}.

The modified neural belief propagation predicts the probability that each key bit was $1$. We denote these probabilities by $o = o_1,..,o_n$. Based on $l$ and $o$, we define the following SAT instance, as described in detail in Sec.~\ref{section:SAT_SOLVER}:
\begin{enumerate}
    \item Define n variables, one per bit $v_1 ,..,v_n$.
    \item $ \psi_{nn}$ a CNF that is induced by the neural belief propagation predictions.
    \item $ \psi_{memory}$ a CNF that is induced by the corrupted key
    \item $ \psi_{AES} $ a CNF that is equivalent to the key expansion constraints.
\end{enumerate}

We run the WBO solver on this instance. The output of our model is the assignment that is returned from the solver.

We note that in contrast to previous cold boot attack methods, the input of our method is a floating vector over $[0,1]$ instead of binary input ${0,1}$. In this way, one can better express the decay model of the memory. In practice, this input can be measured according to the voltage, the memory sector or, the amount of time that elapsed from shutting down the power to the time that the bits were read. However, to compare with previous work on common grounds, our experiments focus entirely on the binary case.

\section{Experiments}
\label{section:results}
We trained our proposed architecture with two types of DRAM memory corruption processes: (i) the theoretical model, where $\delta_1=0$, and (ii) a more realistic model, where $\delta_1=0.001$. For each model, we test with different corruption rates $\delta_0\in [0.40,0.72]$ for the theoretical model and $\delta_0\in [0.50,0.85]$ for the realistic model. 

The training set contains generated random AES $256$ keys. Each batch contains multiple values of corruption rate which are chosen randomly with a uniform distribution at a range of $[{\delta_0}/{4}$,${1-(1-\delta_0)/4}]$ where $\delta_0$ is the corruption rate for the test set in a given experiment. 

During training, we use a batch size of $4$, a learning rate of $1e-4$, and an Adam optimizer \cite{kingma2014adam}. The number of iterations for the neural belief propagation was $L=3$. The parameters $n_l$ were 30 and $n_h$ was $0$, we multiplied the input $l$ by a constant scaling factor of $0.12$.

The S-box network $S_{nn}$ is trained with the Adam optimizer with a batch size of $32$ and a learning rate of $1e-3$. The training set contains all $256$ possible inputs. Our goal is to approximate the Rijndael S-box function on continuous inputs, where it is not defined. Where it is defined, it achieves $100\%$ accuracy and approximates the Rijndael S-box function perfectly.

The baseline methods we compare to include: (1) \cite{tsow2009improved} which recovers AES key schedules from decayed memory images, (2) \cite{tanigaki2015maximum} which is based on a maximum likelihood approach that recovers the key in an imperfect asymmetric decay model, (3) \cite{SATforCBA} which was the first to encode the problem as SAT instant. (4) \cite{Liaobbb} A baseline method that is based on the same SAT solver, but in each instance, we ignore the neural network approximation $o$, as expressed by the formula $\psi_{nn}$. (5) A baseline method, as described in Eq.~\ref{eq:define_H''}, does not employ the S-box.

We run all SAT solvers with a timeout of one hour, with 600 keys per experiment. To reduce the variance of the results, we use the same keys for both our model and the baselines we run. 

\subsection{Ablation variants}

In order to isolate the influence of the neural S-box component $S_{nn}$ on the performance, we perform an ablation analysis. In the first ablation study, we do not use $S_{nn}$, and connect $l$ to the original (without any modifications) belief propagation neural network directly. In these settings, we ignore the non-linear constraints, and use $H''$, a sub-matrix of $H$:  
$$
\forall i,j :  0\leq i < n-k, 0\leq j \leq n
$$
\begin{equation}
\label{eq:define_H''}
H''(i,j)= \begin{cases}
			1, \text{if $i \% b \geq 32$ and j = i }\\
			1, \text{if $i \% b \geq 32$ and j = i+k }\\
			1, \text{if $i \% b \geq 32$ and j = i+k-32 }\\
            0,  \text{otherwise}\\
		 \end{cases}
\end{equation}
\begin{equation}
\label{eq:H''_0}
H l = 0
\end{equation}
This ablation uses only linear constraints. Therefore, we call it as "LC". The second ablation uses $H$, the full matrix, but does not contain neural S-box layers inside the neural belief propagation network.

This ablation uses the original belief propagation neural network architecture, which we denote as "OBPNN".

In the ablation experiments, a neural belief propagation network is constructed, as described in Sec.~\ref{subsection:BP} based on the $H''$ of Eq.~\ref{eq:define_H''}.

\subsection{Results for Theoretical Model $\delta_1=0$}
Tab.~\ref{tab:tsow} presents the results of the theoretical model. As can be seen, our method can handle a corruption rate as high as 72\%, while the method of \citet{tsow2009improved} cannot handle 70\% (and maybe even fails earlier). 

For the lower corruption rate $\delta_0$, we can see that the results are close to $100\%$ for the SAT solver that does not employ the network, for the network without the S-box component followed by the SAT solver. However, when increasing the corruption rate, the difference between the methods becomes more pronounced. The addition of the S-box slightly improves in this setting.

\subsection{Results for Realistic Model $\delta_1=0.001$}
 Tab.~\ref{tab:MLKRA} depicts the results for the realistic model, where $\delta_0=0.001$. Evidently, the baseline method struggles to maintain a high level of performance as the corruption rate increases. Our method, including the two simplified variants, maintains a high performance until a corruption rate of $65\%$, after which the recovery rate starts to drop. For high corruption rates, the advantage of using the network with the S-box becomes clearer.
 
\subsection{Model Analysis}
\begin{figure}[t]
  \centering
      \includegraphics[page=1,width=.5\textwidth]{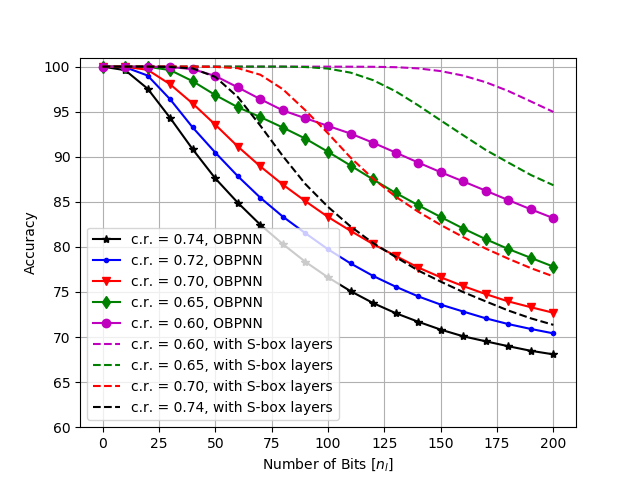}
    \caption{Each line represents the performance per architecture and specific corruption rate (c.r.). The dashed lines represent our architecture and the other lines represent our OBPNN ablation. For each value of $n_l$, we show the accuracy of the bits in $o'$. When this parameter is too high, the probability for more than a few errors increases. On the other hand, if it too low, our network does not influence the search of the SAT solver. }
 \label{fig:convergence}
\end{figure}

In Figure \ref{fig:convergence}, we present (i) the trade-off is expressed by the size of $o'$. As the set size increases, the probability that more than a few errors occur increases as well. We choose $n_l$, with a high probability (for example $99.5\%$) that there are no more than two bits errors in $o'$. Therefore, according to binomial calculation and the figure, we determine that $n_h = 0 $ and $n_l = |o'|$ be in $[20,50]$ for most of the values of $\delta_0$ and $\delta_1$ and.  (i) How the insertion of the S-box layers into the neural belief propagation improves the network performance. Specifically, for each corruption rate, the architecture that contains S-box layers increases the number of bits that can be predicted with high confidence.

\begin{table*}[]
\centering
\caption{Performance evaluation for theoretical model ($\delta_1=0$). The success rate of a cold boot attack for AES-256 with different corruption rates. Higher is better.} % easy model
\label{tab:tsow}
\begin{tabular}{lllllllll}
    \toprule
    Model / Corruption rate & 40\%  & 50\%  & 60\%    & 65\%    & 68\%    & 70\%    & 72\%  & 74\%    \\
    \midrule 
    \cite{tsow2009improved}      & 100.0 & 100.0 & 100.0   & 100.0       &   N/A     & 0.0     & 0.0 & N/A     \\
    \midrule
    MAX-SAT                  & 100.0 & 100.0 & 97.92 & 93.95 & 84.12 & 73.56 & 49.53 &15.95   \\
    Ours LC                  & 100.0 & 100.0 & 99.11 & 95.74 & 88.88 & 81.25 & 53.45 & 18.61  \\
    Ours OBPNN               & 100.0 & 100.0 & 99.43 & 96.41 & 90.52 & 82.27 & 53.90 & 20.27  \\
    Ours                     & 100.0 & 100.0 & 99.51 & 97.05 & 91.20 & 84.10 & 54.52 & 22.35  \\
    \bottomrule 
\end{tabular}
\end{table*}

\begin{table*}[]
\centering
\caption{Performance evaluation for realistic model ($\delta_1=0.001$). The success rate of a cold boot attack for AES-256 with different corruption rates. Higher is better.} % HARD MODEL
\label{tab:MLKRA}
\begin{tabular}{lllllllll}
\toprule
Model / Corruption rate & 50\%  & 55\% & 60\%  & 65\%  & 70\%   & 75\%  \\
\midrule 
\cite{tanigaki2015maximum} (L=1024) & 73.3 & 52.00   & 29.80 & 10.50  & 1.30   & 0.20     \\
\cite{tanigaki2015maximum} (L=2048) & 82.0   & 61.70 & 38.50 & 18.20  & 3.00     & 0.20     \\
\cite{tanigaki2015maximum} (L=4096) & 88.0   & 73.20 & 51.70 & 21.80  & 5.80   & 0.00       \\
\cite{tanigaki2015maximum} (Best)   & 88.0   & 73.20 & 51.70 & 21.80  & 5.80   & 0.20      \\
\midrule
MAX-SAT         & 100.0  & 100.0  & 97.71 & 91.25 & 60.51 & 9.36       \\
Ours LC    & 100.0  & 100.0  & 98.09 & 93.75 & 64.56 & 10.23     \\
Ours OBPNN     & 100.0  & 100.0  & 98.83 & 95.11 & 66.67 & 13.69       \\
Ours     & 100.0  & 100.0  & 99.34 & 96.0 & 66.84 & 14.34       \\
\bottomrule 
\end{tabular}
\end{table*}

\section{Conclusions}
ML is often considered unsuitable for problems in cryptography, due to the combinatorial nature of such problems and the uniform prior of the keys. In this paper, we present convincing evidence in support of employing deep learning in this domain. Specifically, we present a novel method that combines a deep belief propagation neural network and an adapted SAT solver to achieve the state of the art results in the key recovery problem for cold boot attack. 
Our method can recover keys with a high success rate in corruption regions, in which no other method is successful on the AES-256 cipher. 
Our method includes three new techniques: (1) We successfully approximate the S-box transformation by a neural network, despite it being highly non-linear, not differentiable, designed to be resistant to side-channel attacks, and known to be incompatible with a computational graph representation. (2) A new error correcting code representation of the AES family of codes that, unlike previous work, is explicit and also considers all the bits of the original key at once. This approach can be extended to other ciphers, such as Serpent ~\cite{SERPENT}. (3) We are the first, as far as we can ascertain, to combine the approach of the error correcting codes with the SAT solver approach. 
As is shown in our experiments, the hybrid solution we present can to correct bits whose initial value is one but their corrupt value is zero. Detecting this event is challenging, since its prior probability is very low.
The improved success rate of our method on this very popular cipher may have far-reaching implications. 
In addition, the techniques we developed could facilitate an improved success rate in other side channel attacks, for example power analysis attacks \cite{kocherPower}, timing attacks \cite{Timing_attack_Kocher}, and electromagnetic attacks \cite{survey_of_em_attacks,em_first_attack}. 

An interesting direction for future research is to apply more recent neural error correcting code decoders, such as those based on hypernetworks~\cite{nachmani2019hyper}, and evaluate if their improved performance on error correcting codes carries over to the cold boot attack problem.

\section*{Acknowledgement} 
LW thanks the Blavatnik Interdisciplinary Cyber Research Center at Tel-Aviv University, for financial support. The contribution of Eliya Nachmani is part of a Ph.D. thesis research conducted at Tel Aviv University.

% \newpage
\clearpage
\clearpage

\bibliography{example_paper}

\begin{thebibliography}{50}
\providecommand{\natexlab}[1]{#1}
\providecommand{\url}[1]{\texttt{#1}}
\expandafter\ifx\csname urlstyle\endcsname\relax
  \providecommand{\doi}[1]{doi: #1}\else
  \providecommand{\doi}{doi: \begingroup \urlstyle{rm}\Url}\fi

\bibitem[CBT({\natexlab{a}})]{CBT_on_android}
{Cold boot attack affects for android smartphones} kernel description.
\newblock
  \href{https://www.zdnet.com/article/new-cold-boot-attack-affects-seven-years-of-lg-android-smartphones}{link},
  {\natexlab{a}}.
\newblock Accessed: 2010-09-30.

\bibitem[CBT({\natexlab{b}})]{CBT_on_cellphones}
{Cold Boot Attack On CellPhones,} cba for cellphones.
\newblock
  \href{https://www.zdnet.com/article/new-cold-boot-attack-affects-seven-years-of-lg-android-smartphones}{link},
  {\natexlab{b}}.
\newblock Accessed: 2010-09-30.

\bibitem[CBT({\natexlab{c}})]{CBT_on_laptop1}
{Cold Boot Attack On CellPhones,} cba for all laptops.
\newblock
  \href{https://threatpost.com/researchers-heat-up-cold-boot-attack-that-works-on-all-laptops/137466/}{link},
  {\natexlab{c}}.
\newblock Accessed: 2010-09-30.

\bibitem[CBT({\natexlab{d}})]{CBT_on_laptop2}
{Cold Boot Attack On CellPhones,} cba for modern computers.
\newblock
  \href{https://www.zdnet.com/article/new-cold-boot-attack-affects-nearly-all-modern-computers/}{link},
  {\natexlab{d}}.
\newblock Accessed: 2010-09-30.

\bibitem[wbo()]{wbo_solver}
{WBO Partial MAX SAT Solver}.
\newblock \url{http://sat.inesc-id.pt/wbo/}.

\bibitem[Albrecht \& Cid(2011)Albrecht and Cid]{CBaPoliynomialSYS}
Albrecht, M. and Cid, C.
\newblock Cold boot key recovery by solving polynomial systems with noise.
\newblock In \emph{International Conference on Applied Cryptography and Network
  Security}, pp.\  57--72. Springer, 2011.

\bibitem[An et~al.(2020)An, Liang, and Zhang]{an2020high}
An, X., Liang, Y., and Zhang, W.
\newblock High-efficient reed-solomon decoder based on deep learning.
\newblock In \emph{2020 IEEE International Symposium on Circuits and Systems
  (ISCAS)}, pp.\  1--5. IEEE, 2020.

\bibitem[Anderson et~al.(1998)Anderson, Biham, and Knudsen]{SERPENT}
Anderson, R., Biham, E., and Knudsen, L.
\newblock Serpent: A proposal for the advanced encryption standard.
\newblock \emph{NIST AES Proposal}, 174:\penalty0 1--23, 1998.

\bibitem[Bauer et~al.(2016)Bauer, Gruhn, and Freiling]{DDR3_REM}
Bauer, J., Gruhn, M., and Freiling, F.~C.
\newblock Lest we forget: Cold-boot attacks on scrambled ddr3 memory.
\newblock \emph{Digital Investigation}, 16:\penalty0 S65--S74, 2016.

\bibitem[Bayardo~Jr \& Schrag(1997)Bayardo~Jr and Schrag]{CDCL3}
Bayardo~Jr, R.~J. and Schrag, R.
\newblock Using csp look-back techniques to solve real-world sat instances.
\newblock In \emph{Aaai/iaai}, pp.\  203--208. Providence, RI, 1997.

\bibitem[Caciularu \& Burshtein(2020)Caciularu and
  Burshtein]{caciularu2020unsupervised}
Caciularu, A. and Burshtein, D.
\newblock Unsupervised linear and nonlinear channel equalization and decoding
  using variational autoencoders.
\newblock \emph{IEEE Transactions on Cognitive Communications and Networking},
  6\penalty0 (3):\penalty0 1003--1018, 2020.

\bibitem[Daemen \& Rijmen(1999)Daemen and Rijmen]{AES}
Daemen, J. and Rijmen, V.
\newblock Aes proposal: Rijndael.
\newblock 1999.

\bibitem[Doan et~al.(2018)Doan, Hashemi, and Gross]{doan2018neural}
Doan, N., Hashemi, S.~A., and Gross, W.~J.
\newblock Neural successive cancellation decoding of polar codes.
\newblock In \emph{2018 IEEE 19th international workshop on signal processing
  advances in wireless communications (SPAWC)}, pp.\  1--5. IEEE, 2018.

\bibitem[Gross et~al.(2020)Gross, Doan, Ngomseu~Mambou, and
  Ali~Hashemi]{gross2020deep}
Gross, W.~J., Doan, N., Ngomseu~Mambou, E., and Ali~Hashemi, S.
\newblock Deep learning techniques for decoding polar codes.
\newblock \emph{Machine Learning for Future Wireless Communications}, pp.\
  287--301, 2020.

\bibitem[Gruber et~al.(2017)Gruber, Cammerer, Hoydis, and ten
  Brink]{gruber2017deep}
Gruber, T., Cammerer, S., Hoydis, J., and ten Brink, S.
\newblock On deep learning-based channel decoding.
\newblock In \emph{2017 51st Annual Conference on Information Sciences and
  Systems (CISS)}, pp.\  1--6. IEEE, 2017.

\bibitem[Gutmann(2001)]{data_remance_semicondcutr}
Gutmann, P.
\newblock Data remanence in semiconductor devices.
\newblock In \emph{USENIX Security Symposium}, pp.\  39--54, 2001.

\bibitem[Habib et~al.(2020)Habib, Beemer, and Kliewer]{habib2020learning}
Habib, S., Beemer, A., and Kliewer, J.
\newblock Learning to decode: Reinforcement learning for decoding of sparse
  graph-based channel codes.
\newblock \emph{arXiv preprint arXiv:2010.05637}, 2020.

\bibitem[Halderman et~al.(2009)Halderman, Schoen, Heninger, Clarkson, Paul,
  Calandrino, Feldman, Appelbaum, and Felten]{halderman_cba}
Halderman, J.~A., Schoen, S.~D., Heninger, N., Clarkson, W., Paul, W.,
  Calandrino, J.~A., Feldman, A.~J., Appelbaum, J., and Felten, E.~W.
\newblock Lest we remember: cold-boot attacks on encryption keys.
\newblock \emph{Communications of the ACM}, 52\penalty0 (5):\penalty0 91--98,
  2009.

\bibitem[Jiang et~al.(2019)Jiang, Kim, Asnani, Kannan, Oh, and
  Viswanath]{jiang2019turbo}
Jiang, Y., Kim, H., Asnani, H., Kannan, S., Oh, S., and Viswanath, P.
\newblock Turbo autoencoder: Deep learning based channel codes for
  point-to-point communication channels.
\newblock In \emph{Advances in Neural Information Processing Systems}, pp.\
  2754--2764, 2019.

\bibitem[Kamal \& Youssef(2010)Kamal and Youssef]{SATforCBA}
Kamal, A.~A. and Youssef, A.~M.
\newblock Applications of sat solvers to aes key recovery from decayed key
  schedule images.
\newblock In \emph{2010 Fourth International Conference on Emerging Security
  Information, Systems and Technologies}, pp.\  216--220. IEEE, 2010.

\bibitem[Kamassury \& Silva(2020)Kamassury and Silva]{kamassury2020iterative}
Kamassury, J. K.~S. and Silva, D.
\newblock Iterative error decimation for syndrome-based neural network
  decoders.
\newblock \emph{arXiv preprint arXiv:2012.00089}, 2020.

\bibitem[Kim et~al.(2018{\natexlab{a}})Kim, Jiang, Kannan, Oh, and
  Viswanath]{kim2018deepcode}
Kim, H., Jiang, Y., Kannan, S., Oh, S., and Viswanath, P.
\newblock Deepcode: Feedback codes via deep learning.
\newblock In \emph{Advances in Neural Information Processing Systems (NIPS)},
  pp.\  9436--9446, 2018{\natexlab{a}}.

\bibitem[Kim et~al.(2018{\natexlab{b}})Kim, Jiang, Rana, Kannan, Oh, and
  Viswanath]{kim2018communication}
Kim, H., Jiang, Y., Rana, R., Kannan, S., Oh, S., and Viswanath, P.
\newblock Communication algorithms via deep learning.
\newblock \emph{arXiv preprint arXiv:1805.09317}, 2018{\natexlab{b}}.

\bibitem[Kingma \& Ba(2014)Kingma and Ba]{kingma2014adam}
Kingma, D.~P. and Ba, J.
\newblock Adam: A method for stochastic optimization.
\newblock \emph{arXiv preprint arXiv:1412.6980}, 2014.

\bibitem[Kocher et~al.(1998)Kocher, Jaffe, Jun, et~al.]{kocherPower}
Kocher, P., Jaffe, J., Jun, B., et~al.
\newblock Introduction to differential power analysis and related attacks,
  1998.

\bibitem[Kocher(1996)]{Timing_attack_Kocher}
Kocher, P.~C.
\newblock Timing attacks on implementations of diffie-hellman, rsa, dss, and
  other systems.
\newblock In Koblitz, N. (ed.), \emph{Advances in Cryptology --- CRYPTO '96},
  pp.\  104--113, Berlin, Heidelberg, 1996. Springer Berlin Heidelberg.
\newblock ISBN 978-3-540-68697-2.

\bibitem[Liao et~al.(2013)Liao, Zhang, Koshimura, Fujita, and
  Hasegawa]{Liaobbb}
Liao, X., Zhang, H., Koshimura, M., Fujita, H., and Hasegawa, R.
\newblock Using maxsat to correct errors in aes key schedule images.
\newblock In \emph{2013 IEEE 25th international conference on tools with
  artificial intelligence}, pp.\  284--291. IEEE, 2013.

\bibitem[Manquinho et~al.(2009)Manquinho, Silva, and Planes]{WBO3}
Manquinho, V.~M., Silva, J. P.~M., and Planes, J.
\newblock Algorithms for weighted boolean optimization.
\newblock In Kullmann, O. (ed.), \emph{Theory and Applications of
  Satisfiability Testing - {SAT} 2009, 12th International Conference, {SAT}
  2009, Swansea, UK, June 30 - July 3, 2009. Proceedings}, volume 5584 of
  \emph{Lecture Notes in Computer Science}, pp.\  495--508. Springer, 2009.
\newblock \doi{10.1007/978-3-642-02777-2\_45}.
\newblock URL \url{https://doi.org/10.1007/978-3-642-02777-2\_45}.

\bibitem[Manquinho et~al.(2010)Manquinho, Martins, and Lynce]{WBO2}
Manquinho, V.~M., Martins, R., and Lynce, I.
\newblock Improving unsatisfiability-based algorithms for boolean optimization.
\newblock In Strichman, O. and Szeider, S. (eds.), \emph{Theory and
  Applications of Satisfiability Testing - {SAT} 2010, 13th International
  Conference, {SAT} 2010, Edinburgh, UK, July 11-14, 2010. Proceedings}, volume
  6175 of \emph{Lecture Notes in Computer Science}, pp.\  181--193. Springer,
  2010.
\newblock \doi{10.1007/978-3-642-14186-7\_16}.
\newblock URL \url{https://doi.org/10.1007/978-3-642-14186-7\_16}.

\bibitem[Marques-Silva \& Sakallah(1999)Marques-Silva and Sakallah]{CDCL2}
Marques-Silva, J.~P. and Sakallah, K.~A.
\newblock Grasp: A search algorithm for propositional satisfiability.
\newblock \emph{IEEE Transactions on Computers}, 48\penalty0 (5):\penalty0
  506--521, 1999.

\bibitem[Martins et~al.(2011)Martins, Manquinho, and Lynce]{WBO1}
Martins, R., Manquinho, V.~M., and Lynce, I.
\newblock Exploiting cardinality encodings in parallel maximum satisfiability.
\newblock In \emph{{IEEE} 23rd International Conference on Tools with
  Artificial Intelligence, {ICTAI} 2011, Boca Raton, FL, USA, November 7-9,
  2011}, pp.\  313--320. {IEEE} Computer Society, 2011.
\newblock \doi{10.1109/ICTAI.2011.54}.
\newblock URL \url{https://doi.org/10.1109/ICTAI.2011.54}.

\bibitem[Martins et~al.(2012)Martins, Manquinho, and Lynce]{UNSAT}
Martins, R., Manquinho, V.~M., and Lynce, I.
\newblock On partitioning for maximum satisfiability.
\newblock In Raedt, L.~D., Bessiere, C., Dubois, D., Doherty, P., Frasconi, P.,
  Heintz, F., and Lucas, P. J.~F. (eds.), \emph{{ECAI} 2012 - 20th European
  Conference on Artificial Intelligence. Including Prestigious Applications of
  Artificial Intelligence {(PAIS-2012)} System Demonstrations Track,
  Montpellier, France, August 27-31 , 2012}, volume 242 of \emph{Frontiers in
  Artificial Intelligence and Applications}, pp.\  913--914. {IOS} Press, 2012.
\newblock \doi{10.3233/978-1-61499-098-7-913}.
\newblock URL \url{https://doi.org/10.3233/978-1-61499-098-7-913}.

\bibitem[M{\"u}ller et~al.(2012)M{\"u}ller, Spreitzenbarth, and
  Freiling]{CBT_on_cellphones2}
M{\"u}ller, T., Spreitzenbarth, M., and Freiling, F.~C.
\newblock Forensic recovery of scrambled telephones, 2012.

\bibitem[Nachmani \& Wolf(2019)Nachmani and Wolf]{nachmani2019hyper}
Nachmani, E. and Wolf, L.
\newblock Hyper-graph-network decoders for block codes.
\newblock \emph{Advances in Neural Information Processing Systems},
  32:\penalty0 2329--2339, 2019.

\bibitem[Nachmani et~al.(2016)Nachmani, Be'ery, and
  Burshtein]{nachmani2016learning}
Nachmani, E., Be'ery, Y., and Burshtein, D.
\newblock Learning to decode linear codes using deep learning.
\newblock In \emph{2016 54th Annual Allerton Conference on Communication,
  Control, and Computing (Allerton)}, pp.\  341--346. IEEE, 2016.

\bibitem[Nyberg(1991)]{nyberg1991perfect}
Nyberg, K.
\newblock Perfect nonlinear s-boxes.
\newblock In \emph{Workshop on the Theory and Application of of Cryptographic
  Techniques}, pp.\  378--386. Springer, 1991.

\bibitem[Ooi \& Kam(2009)Ooi and Kam]{DefCBA}
Ooi, J.~G. and Kam, K.~H.
\newblock A proof of concept on defending cold boot attack.
\newblock In \emph{2009 1st Asia Symposium on Quality Electronic Design}, pp.\
  330--335. IEEE, 2009.

\bibitem[Quisquater(2000)]{em_first_attack}
Quisquater, J.-J.
\newblock A new tool for non-intrusive analysis of smart cards based on
  electro-magnetic emissions. the sema and dema methods.
\newblock \emph{Eurocrypt2000 rump session}, 2000.

\bibitem[Raviv et~al.(2020)Raviv, Raviv, and Be’ery]{raviv2020data}
Raviv, T., Raviv, N., and Be’ery, Y.
\newblock Data-driven ensembles for deep and hard-decision hybrid decoding.
\newblock In \emph{2020 IEEE International Symposium on Information Theory
  (ISIT)}, pp.\  321--326. IEEE, 2020.

\bibitem[Sayakkara et~al.(2019)Sayakkara, Le-Khac, and
  Scanlon]{survey_of_em_attacks}
Sayakkara, A., Le-Khac, N.-A., and Scanlon, M.
\newblock A survey of electromagnetic side-channel attacks and discussion on
  their case-progressing potential for digital forensics.
\newblock \emph{Digital Investigation}, 29:\penalty0 43--54, 2019.

\bibitem[Shlezinger et~al.(2020)Shlezinger, Fu, and
  Eldar]{shlezinger2020deepsic}
Shlezinger, N., Fu, R., and Eldar, Y.~C.
\newblock Deepsic: Deep soft interference cancellation for multiuser mimo
  detection.
\newblock \emph{IEEE Transactions on Wireless Communications}, 2020.

\bibitem[Silva \& Sakallah(2003)Silva and Sakallah]{CDCL1}
Silva, J. P.~M. and Sakallah, K.~A.
\newblock Grasp—a new search algorithm for satisfiability.
\newblock In \emph{The Best of ICCAD}, pp.\  73--89. Springer, 2003.

\bibitem[Skorobogatov(2005)]{Data_remanence_flash}
Skorobogatov, S.
\newblock Data remanence in flash memory devices.
\newblock In \emph{International Workshop on Cryptographic Hardware and
  Embedded Systems}, pp.\  339--353. Springer, 2005.

\bibitem[Tal \& Vardy(2013)Tal and Vardy]{tal2013construct}
Tal, I. and Vardy, A.
\newblock How to construct polar codes.
\newblock \emph{IEEE Transactions on Information Theory}, 59\penalty0
  (10):\penalty0 6562--6582, 2013.

\bibitem[Tanigaki \& Kunihiro(2015)Tanigaki and Kunihiro]{tanigaki2015maximum}
Tanigaki, T. and Kunihiro, N.
\newblock Maximum likelihood-based key recovery algorithm from decayed key
  schedules.
\newblock In \emph{ICISC 2015}, pp.\  314--328. Springer, 2015.

\bibitem[Teng et~al.(2019)Teng, Wu, Ho, and Wu]{teng2019low}
Teng, C.-F., Wu, C.-H.~D., Ho, A. K.-S., and Wu, A.-Y.~A.
\newblock Low-complexity recurrent neural network-based polar decoder with
  weight quantization mechanism.
\newblock In \emph{ICASSP 2019-2019 IEEE International Conference on Acoustics,
  Speech and Signal Processing (ICASSP)}, pp.\  1413--1417. IEEE, 2019.

\bibitem[Tsow(2009)]{tsow2009improved}
Tsow, A.
\newblock An improved recovery algorithm for decayed aes key schedule images.
\newblock In \emph{International Workshop on Selected Areas in Cryptography},
  pp.\  215--230. Springer, 2009.

\bibitem[Xu et~al.(2017)Xu, Wu, Ueng, You, and Zhang]{xu2017improved}
Xu, W., Wu, Z., Ueng, Y.-L., You, X., and Zhang, C.
\newblock Improved polar decoder based on deep learning.
\newblock In \emph{2017 IEEE International workshop on signal processing
  systems (SiPS)}, pp.\  1--6. IEEE, 2017.

\bibitem[Yitbarek et~al.(2017)Yitbarek, Aga, Das, and Austin]{DDR_and_MODERN}
Yitbarek, S.~F., Aga, M.~T., Das, R., and Austin, T.
\newblock Cold boot attacks are still hot: Security analysis of memory
  scramblers in modern processors.
\newblock In \emph{2017 IEEE International Symposium on High Performance
  Computer Architecture (HPCA)}, pp.\  313--324. IEEE, 2017.

\bibitem[Zengler()]{LogicNGgithub}
Zengler, C.
\newblock Logicng library.
\newblock \url{https://github.com/logic-ng/LogicNG}.

\end{thebibliography}
\bibliographystyle{icml2021}

\end{document}